\documentclass[aps,prb,twocolumn,amsmath,amssymb,superscriptaddress]{revtex4}
\usepackage{amssymb}
\usepackage{amsmath}
\usepackage{graphicx}
\usepackage{subfigure}
\usepackage{textcomp}
\usepackage{color}
\usepackage{amsfonts}
\usepackage{bbold}
\usepackage{epsfig}
\usepackage{mathrsfs}
\usepackage{hyperref}

\begin{document}
\title{Oxygen doping and polaron magnetic coupling in Alq$_3$ films}
\author{Andrea Droghetti}
\affiliation{Nano-Bio Spectroscopy Group and European Theoretical Spectroscopy Facility, Materials Physics Center, University of the Basque Country, Av.Tolosa 72 ,20018 San Sebastian, Spain}
\affiliation{School of Physics and CRANN Institute, Trinity College, Dublin 2, Ireland}

\begin{abstract}
The understanding of the Physics underlying the performances of spin-valve devices comprising organic semiconductors is still incomplete. 
According to some recent models, spin transport takes place in an impurity band inside the fundamental gap of the organic semiconductor. 
This seems to be confirmed by recent experiments performed with La$_{0.7}$Sr$_{0.3}$MnO$_3$/Alq$_3$/AlO$_x$/Co devices. 
The reported results suggest a possible correlation between the magnetoresistance and the variable oxygen doping in the Alq$_3$ spacer. 
In this paper we investigate the electronic and magnetic properties of O$_2$ molecules and ions in Alq$_3$ films by means of first-principles calculations 
to establish whether oxygen plays any important role for spin transport in La$_{0.7}$Sr$_{0.3}$MnO$_3$/Alq$_3$/AlO$_x$/Co devices. 
The conclusion is that it does not. In fact, we show that O$_2$ molecules do not form an impurity band and there is no magnetic interaction between them. In contrast, we suggest that spin-transport may be enabled by the direct exchange coupling between Alq$_3^-$ ions. 
\end{abstract}

\maketitle

\section{Introduction}

More than a decade ago organic semiconductors (OSCs) were proposed as an ideal medium for spin-transport \cite{Alek, Alek_review} owing to their long spin-relaxation time \cite{Szulczewski}. 
%of their low spin-orbit coupling and hyperfine interaction, which result in a long spin-relaxation time\cite{Szulczewski}. 
%Since then, there have been many experiments, which have succeded in measuring magnetoresistance (MR) in hybrid spin-valve devices comprising an OSC film in between two ferromagnetic electrodes\cite{Xiong,Dediu,DSun,Lin,Gobbi,XSun1}, 
%but there have also been several other experiments, which have failed\cite{Jiang}. 
%Since then many experiments have succeeded in measuring magnetoresistance (MR) in hybrid spin-valve devices comprising an OSC film
%between two ferromagnetic electrodes\cite{Xiong,Dediu,DSun,Lin,Gobbi,XSun1}. At the same time, however, there have been also a number of experiments, which have failed\cite{Lin,Jiang}. 
Since then a large number of experiments have investigated the performances and the underlying physics of the so-called organic spin-valve devices, which comprise an OSC spacer in between two ferromagnetic electrodes \cite{Devkota}. Some of these experiments 
have succeeded in measuring magnetoresistance (MR) \cite{Xiong,Dediu,DSun,Lin,Gobbi,XSun1}, but, at the same time, some others have failed \cite{Lin,Jiang}. 
Overall these seemingly contradictory reports 
%have been very difficult to rationalize and 
 show that our understanding of spin-injection and transport in OSCs is still largely incomplete.  \\
A few studies unequivocally proved that the injection of spin-polarized charge carriers from ferromagnetic metals into OSCs could be achieved either optically \cite{Mirko, Mirko2} or electrically \cite{Nguyen,Ruiz}.
The injection potential barriers were generally estimated 
of the order of 1~V~\cite{XSun1,Droghetti,Gobbi2}. 
These large values however contrast the results of most spin-valve experiments, where the MR was detected only at low bias voltages ($\sim 0.1$~V)~\cite{Alek_review,Xiong,Hueso}. 
Because of this issue one can question whether the measured MR is really due to spin transport over molecular orbitals. 
Furthermore, we note that the Hanle effect, which is commonly considered as the ultimate proof of spin-injection, was reported absent \cite{Riminucci, Grunewald}, 
while most devices showing MR were surprisingly very conductive \cite{Riminucci2} with their performances not disrupted by the conductivity mismatch \cite{Schmidt}.\\ 
%This body of experimental observations has been partly rationalized by using an impurity band model firstly introduced by Yu\cite{Yu1,Yu2}. Transport through an OSC film in a hybrid spin-valves takes place in an impurity band. 
%Some of these impurities hosts a unpaired carrier and therefore a magnetic moment. At sufficiently high impurity concentration charge current is due to the hopping of these carriers, while spin current is enabled by the exchange coupling between the magnetic moments, which can be direct\cite{Yu2} or indirect\cite{Droghetti_model} according to different variations of the model. Charge and spin currents propagate through different regions of the film according to a two-fluid picture\cite{Droghetti_model}.\\
This body of experimental observations has been partly rationalized by using an impurity band model firstly introduced by Yu \cite{Yu1,Yu2}. 
%Transport through an OSC film in a hybrid spin-valves takes place in an impurity band. The presence of the impurity band indicates that charge carriers already exist before the application of a bias voltage. 
In an OSC film there are impurities, which introduce charge carriers and which give rise to an impurity band inside the OSC transport gap. 
At sufficiently high impurity concentrations the charge current is due to hopping of the carriers between impurities. 
 Concurrently, carriers can become localized and unpaired at impurity sites, which therefore have a magnetic moment. 
Spin current is then enabled by the exchange coupling between these magnetic moments %, which can be direct\cite{Yu2} or indirect\cite{Droghetti_model} according to different variations of the model,
and is carried by spin-waves. Charge and spin propagate through different regions of the film as two fluids \cite{Droghetti_model} with different diffusion constants \cite{Yu1}. 
The impurity band model can qualitatively explain the typical $I-V$ characteristic curve of spin-valves, the dependence of the MR on the bias voltage, the absence of the conductivity mismatch problem and of the Hanle effect \cite{Yu1,Yu2}.
Furthermore, it has been recently used to calculate the spin-diffusion length in polymers achieving a remarkable agreement with experimental results \cite{Wang}.\\
%Recently La$_{0.7}$Sr$_{0.3}$MnO$_3$ (LSMO)/Alq$_3$/AlO$_x$/Co hybrid spin-valve devices have been studied by Riminucci {\it et al.} to validate the impurity model\cite{Riminucci2}. 
%In these devices both the charge carriers density and the MR can be modulated systematically through multilevel resistive switching\cite{Hueso2, Prezioso, Prezioso2,Grunewald2}. 
%The experimental results link unequivocally the variable oxygen doping in the Alq$_3$ film with impurity driven effects\cite{Bergenti}. 
%The MR is then interpreted as a result of spin transport across oxygen states\cite{Riminucci3}.
%According to the proposed picture\cite{Bergenti, Riminucci3} AlO$_x$ tunnel layers incorporate O$_2^-$ ions that are trapped near positively charged oxygen vacancies\cite{Tan}. 
%In LSMO/Alq$_3$/AlO$_x$/Co spin-valves 
%These ions are then able to migrate into and out the AlO$_x$ layers under electrical bias eventually modifying the properties of the Alq$_3$ film\cite{Bergenti}. O$_2^-$ doping in the Alq$_3$ film correlates with the carriers concentration. At low carriers concentration neither MR nor multi-level resistive switching is observed. 
%Instead, when the carriers concentration is sufficiently high, both effects are measured.\\
During the last few years Riminucci {\it et al.}  have investigated La$_{0.7}$Sr$_{0.3}$MnO$_3$/Alq$_3$/AlO$_x$/Co spin-valve devices to validate the impurity model \cite{Riminucci2}. In these devices both the charge carrier density and the MR can be modulated systematically through multilevel resistive switching \cite{Hueso2, Prezioso, Prezioso2,Grunewald2}. At low carrier concentrations the MR is not observed, while it is measured, when the carrier concentration is sufficiently high ($\sim 10^{19}$ cm$^{-3}$). The authors suggested a possible correlation between this behavior and the variable oxygen doping in the Alq$_3$ spacer \cite{Bergenti}. 
According to the proposed picture \cite{Bergenti, Riminucci3} the AlO$_x$ tunnel layers incorporate O$^-$ ions that are trapped near positively charged oxygen vacancies \cite{Tan}. These ions are then able to migrate out of the AlO$_x$ layers under the application of a large (several V) electrical bias pulse forming O$_2^-$ ions and eventually doping the Alq$_3$ film \cite{Bergenti}.
The MR is then measured at low voltages ($\sim 0.1$ V) and it is interpreted as a result of spin transport across oxygen states in line with the impurity band model \cite{Riminucci3}.\\
In this paper we investigate the electronic and magnetic properties of oxygen molecules O$_2$ and ions O$_2^-$ in Alq$_3$ films by means of first-principles calculations to establish whether they play any important role for charge and spin transport. 
The conclusion is that they probably do not. 
Specifically, we show that O$_2$ molecules do not form an impurity band. Furthermore, no magnetic interaction is found between pairs of either O$_2$ or O$_2^-$. 
In contrast, Alq$_3^-$ ions, which have spin $1/2$, are exchanged coupled via the $\pi-\pi$ interaction of their ligands. This exchange coupling, which is direct, can reach almost room temperature. %We therefore propose that in (LSMO)/Alq$_3$/AlO$_x$/Co hybrid SV devices spin is transported across Alq$_3^-$ ions and not oxygen states as previously suggested\cite{Bergenti,Riminucci3}.  
Hence, we propose that spin transport in La$_{0.7}$Sr$_{0.3}$MnO$_3$/Alq$_3$/AlO$_x$/Co spin-valves does not involve oxygen states as suggested previously by Riminucci {\it et al.} \cite{Riminucci3}, but it may be enabled by the exchange coupling between Alq$_3^-$ ions instead.\\

\begin{figure}
	\centering
		\includegraphics[width=0.53\textwidth,clip=true]{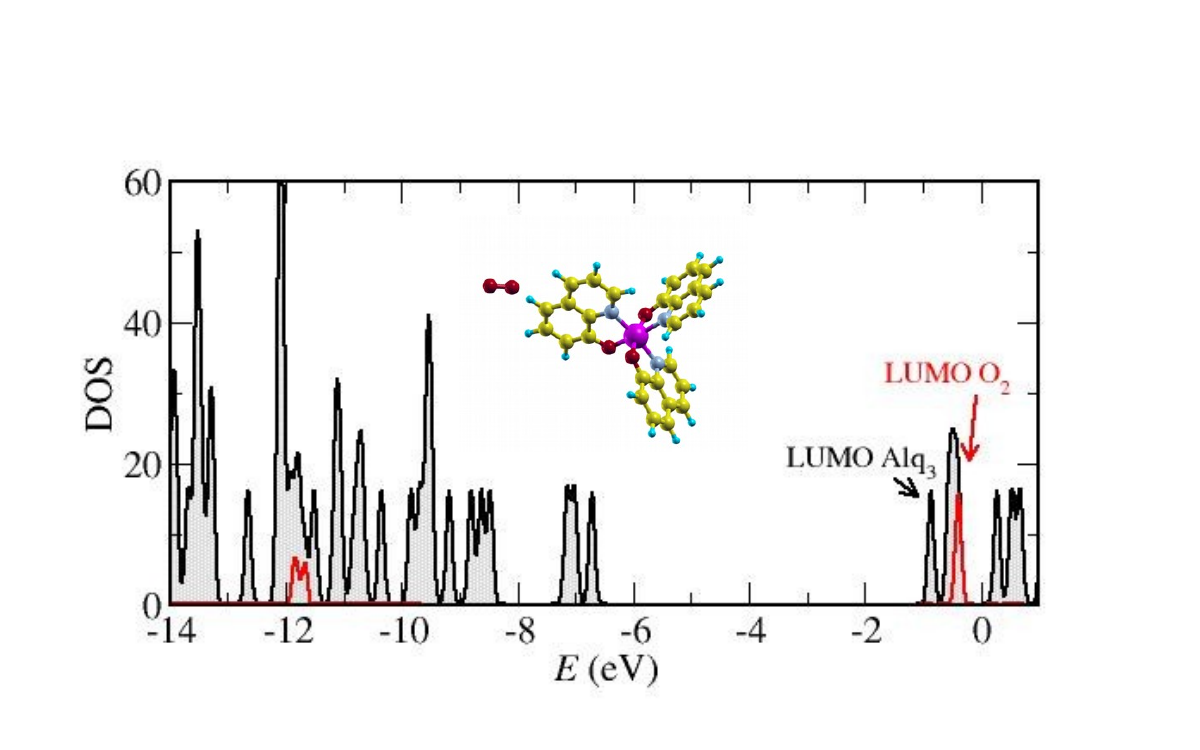}
	\caption{DOS calculated by using G$_0$W$_0$ for a O$_2$-Alq$_3$ complex. The DOS projected over the O$_2$ molecule is in red.}
	\label{fig.dos}
\end{figure}

\section{Results}
\subsection{Electronic properties}
To begin with we investigate whether any of the O$_2$ frontier states lays inside the transport gap of Alq$_3$ eventually giving rise to an impurity band. We therefore consider a cluster comprising an O$_2$ molecule in close proximity to an Alq$_3$ molecule (see inset of Fig. \ref{fig.dos}) and we analyze the relative energy level alignment obtained from first-principles calculations.\\
All presented calculations are performed by using the all-electron code FHI-AIMS \cite{Blum, Ren, Caruso}. 
The atomic positions were optimized by Density Functional Theory (DFT) with the Tkatchenko-Scheffler (TS) van der Waals-corrected method \cite{Tkatchenko}
until the forces were smaller than 0.01 eV/\AA. We consider
different initial configurations for O$_2$ and Alq$_3$ . We generally find that
O$_2$ is unlikely to go close to the Al-center of Alq$_3$ . After optimization O$_2$
always ends up quite far from the Alq$_3$ molecule as, for example, shown
in Fig. 1. For such Alq$_3$-O$_2$ separation the electronic properties of the
cluster are quite independent on the relative position of the two molecules and, additionally, they do not depend on the orientation of O$_2$.
We therefore describe here only the results obtained for the cluster in
Fig. \ref{fig.dos}.\\
The energy level alignment is inferred from the density of states (DOS) computed with the G$_0$W$_0$ approximation of many-body perturbation theory \cite{Ferdi,Ren, Caruso}. It is important to remind that the electronic states in many-body perturbation theory can be rigorously interpreted as the quasiparticle excitations measured in photoemission experiments, unlike the Kohn-Sham states of DFT \cite{Kronik}. 
Many-body perturbation theory therefore represents the appropriate theoretical framework to study energy level alignment in materials.
The initial state for G$_0$W$_0$ is obtained from a spin-polarized DFT calculation with the PBE0 hybrid exchange correlation functional \cite{PBE0} using the same computation details as in Refs. \cite{Marom,DroghettiGW}. Spin-polarized calculations are required to describe O$_2$, which is a magnetic molecule of spin $1$. 
As we showed in a previous work \cite{DroghettiGW}, G$_0$W$_0$ yields excellent results for the occupied spectra of Alq$_3$ compared to experiments. 
Similarly, it is also expected to provide reliable estimates for the energies of unoccupied states, although it is hard in practice to achieve the basis set convergence \cite{Nabok}. 
Because of that, we also use constrained-DFT (c-DFT) \cite{Dederichs, Kaduk, Wu1,Wu2} with the local spin density approximation for the exchange correlation potential as a complementary approach. In fact it has been demonstrated that c-DFT generally returns quite accurate results for the energies of the frontier orbitals of molecules inside crystals and clusters \cite{andrea_cdft,Gaul}. 
Specifically, the energy of the lowest unoccupied molecular orbital (LUMO) of Alq$_3$ (O$_2$) is obtained in c-DFT 
by calculating the opposite of the electron affinity $A$
via the finite-energy difference $\epsilon_{\mathrm{LUMO}}=-A=E_{-}- E$. $E$ is the energy of Alq$_3$ (O$_2$) and $E_{-}$ is the energy 
of the anion Alq$_3^-$ (O$_2^-$) obtained by enforcing the localization on the molecule of the added electron, which, otherwise, would spread over the whole system in a nonphysical way. 
Similarly, the energy of highest occupied molecular orbital (HOMO) is calculated as the opposite of the ionization potential $I$
according to $\epsilon_{\mathrm{HOMO}}=-I=E- E_{+}$ 
with $E_{+}$ the energy of the cation Alq$_3^+$ (O$_2^+$). The c-DFT calculations are performed by using the default basis set referred as ``tight'' in the FHI-AIMS manual. \\
\begin{table}[t]
{%
\begin{tabular}{lccc}\hline\hline
      & $\epsilon_{\mathrm{HOMO}}$ (eV)  & $\epsilon_{\mathrm{LUMO}}$ (eV)    \\  \hline\hline
O$_2$ &   -11.34  &     -0.37
\\ 
Alq$_3$ &  -6.37  &   -1.21 \\
\hline
\end{tabular}}
\caption{HOMO and LUMO energies of Alq$_3$ and O$_2$ calculated by c-DFT.}\label{Tab.cdft}
\end{table}

The G$_0$W$_0$ DOS is presented in Fig. \ref{fig.dos}, where
the HOMO and the LUMO of Alq$_3$ and O$_2$ can be readily recognized
(we note that, since O$_2$ is magnetic, its HOMO and LUMO are respectively spin up and spin down orbitals). The c-DFT estimates are listed in Tab. \ref{Tab.cdft}. 
Overall, the two different methods give results in very good agreement making us confident about the reliability of our predictions. The O$_2$ HOMO is more than $5$ eV below the Alq$_3$ HOMO and the O$_2$ LUMO is almost $1$ eV above the LUMO
of Alq$_3$. This means that, firstly, O$_2$ does not form an impurity band inside the Alq$_3$ transport gap and, secondly, an extra electron added to the whole system will not reduce O$_2$ to O$_2^-$, but it will rather go into the LUMO of Alq$_3$ giving Alq$_3^-$. Hence, no O$_2^-$ ions are expected in Alq$_3$ films unlike what proposed in Refs. \cite{Bergenti, Riminucci3}. In La$_{0.7}$Sr$_{0.3}$MnO$_3$/Alq$_3$/AlO$_x$/Co spin-valves, if O$_2^{-}$ impurities are driven into the Alq$_3$ film from the AlO$_x$ layers, they will donate the electron to an Alq$_3$ molecule. 
This is the first important observation of this paper.\\
The energy level alignment studied so far refers to vertical quasi-particle excitation energies. This means that we neglected the
molecular conformational changes induced by the addition of an electron and which increase the electron affinity. The energy associated to
such ionic effect, called inner-shell re-organization energy \cite{Nelsen} and usually labeled $\lambda$, can be estimated by subtracting the energy of the anion
at the neutral molecule geometry from the energy of the anion at its own relaxed geometry in gas phase \cite{Rudnev} (we assume that the external
reorganization energy is the same for the two molecules and we therefore neglect it). The increased electron affinity, which is called
adiabatic, is given by $A_a = A + \lambda$ . In our calculations we find that $\lambda$ is $0.06$ eV and $0.49$ eV for Alq$_3$ and O$_2$ respectively. 
The reorganization energy is therefore much larger for O$_2$ than for Alq$_3$. 
However $A_a$ is lower for O$_2$ than for Alq$_3$. 
Hence, even including ionic contributions, the picture presented above remains valid and a dopant electron prefers
to reduce Alq$_3$ rather than O$_2$.\\
In addition to O$_2$, we also consider the effect of other dopants by performing similar G$_0$W$_0$ calculations. It is worth to remark that, in this paper, we focus only on the electronic structure, while understanding possible chemical reactions, which are still very much debated \cite{Scholz}, is beyond our goals. We start by considering a single O atom or O$^-$ ion. These establish a covalent bond with Alq$_3$ during the DFT optimization. 
Hence, we can not have free O$^{(-)}$ in Alq$_3$ films. The resulting complex does not for an impurity band and moreover is non-magnetic.
We also introduce a water molecule, which is known to be a common intrinsic impurity in Alq$_3$ films. 
However, we predict that H$_2$O similarly to O$_2$ does not have any state in the transport gap of Alq$_3$ and therefore it can not give rise to an impurity band. %Furthermore water molecules are usually found to react with Alq$_3$, although what the products are is still very much debated in the literature\cite{Scholz}.
Finally, we also propose some triatomic molecules, for example O$_3$ and NO$_2$, as alternative dopants. Since these molecules have a large electron affinity they induce an intra-gap state in Alq$_3$.  For instance, the O$_3$ LUMO is at about 3.2 (3.4) eV below (above) the Alq$_3$ LUMO (HOMO). However, as far as we know, no evidence for the presence of this type of  molecules in Alq$_3$ films has ever been reported and it is not clear what mechanism would be responsible for their incorporation into La$_{0.7}$Sr$_{0.3}$MnO$_3$/Alq$_3$/AlO$_x$/Co spin-valves, in particular because these devices are usually fabricated in ultra-high vacuum condition. This may deserve more experimental investigations in the future.\\
%Besides that, it is worth to remark that in this letter, our focus is only on the electronic structure of these impurities by using G$_0$W$_0$ calculations.  Their possible chemical reactions with Alq$_3$, which are still very much debated\cite{Scholz}, is beyond our goals here and the capabilities of the employed computational methods and they will deserve more theoretical investigations in the future.\\
Although our G$_0$W$_0$ and c-DFT calculations are performed for clusters to limit the computational overhead, the results can be readily extrapolated towards those of a film by adding electrostatic polarization effects\cite{Sato}. 
These will induce only a rigid shift of unoccupied (occupied) states towards lower (higher) energies. As such, they will change the absolute energy-position of the various states, but not their relative position, which is ultimately what we are interested in. 
The polarization-induced shift can be calculated by assuming that a molecule is a charged point embedded
in a dielectric medium \cite{Abramson}. Since an Alq$_3$ film has dielectric constant of $3$ \cite{Knox}, the energy shift is estimated to be about $-(+)0.9$ eV for unoccupied (occupied) states, but, as just stated, this will not change the conclusions drawn in the previous paragraphs.\\
\subsection{Magnetic properties}
\begin{figure}
	\centering
		\includegraphics[width=0.45\textwidth,clip=true]{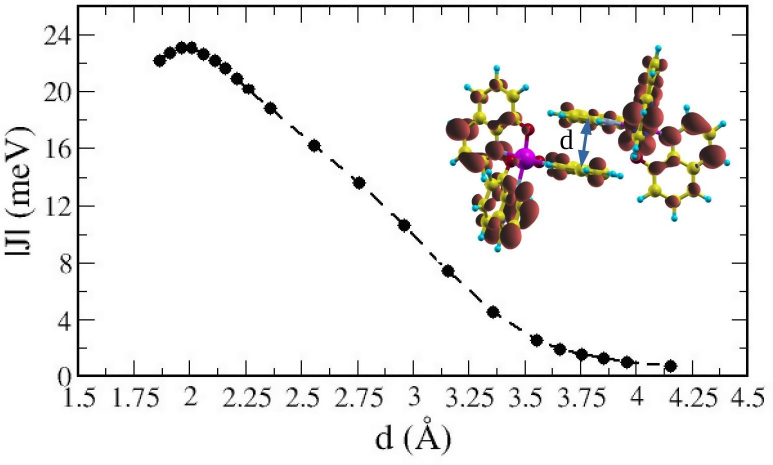}
	\caption{Exchange coupling $J$ as a function of the distance $d$ between the $\pi$-stacked ligands of two Alq$_3^-$ ions. Inset: the magnetization isosurface for the Alq$_3^-$ ions.}
	\label{fig.J}
\end{figure}
After having addressed the electronic structure of the impurities in Alq$_3$, we now examine the magnetic properties. O$_2$ and O$_2^-$ have spin $1$ and $1/2$ respectively and our calculations correctly reproduce that. 
The main question is whether we can have magnetic interaction between pairs of O$_2^{(-)}$ as required to enable spin transport according to the impurity band model. 
To provide an answer we consider several huge clusters, which contain fifteen Alq$_3$
molecules and two O$_2^{(-)}$ impurities each. These clusters differ one from another for the relative positions and distances of the  O$_2^{(-)}$ impurities
allowing us to study the spatial dependence of the eventual magnetic interaction. The atomic positions are fully optimized by using DFT with
the TS method. We then assume that the magnetic interaction between the two impurities can be described through a Heisenberg model with
the exchange coupling $J$, which is computed by means of the broken symmetry approach \cite{Postnikov} as follows. We perform a first calculation with
the magnetic moments of the two impurities constrained to be parallel (P) and we obtain the total energy $E_P$. Afterward, we perform a second
calculation with anti-parallel magnetic moments and we obtain the total energy $E_{AP}$. The exchange coupling $J$ is finally given by
$J = (E_P - E_{AP})/4S$ \cite{Postnikov} with $S$ equal to 1 and 1/2 for O$_2$ and O$_2^-$ respectively. These calculations are repeated for many clusters. 
However, in no case we find a finite $J$ within the accuracy of our calculations (0.1 meV). Hence we conclude that spin transport in Alq3 does not involve
the exchange coupling between oxygen states in contrast to what suggested in Refs. \cite{Bergenti,Riminucci3}.\\
While Alq$_3$ is non-magnetic, the ion Alq$_3^-$ has a spin 1/2 with the added electron that is localized over the pyridyl moieties of the ligands \cite{Tarafder,Curioni, Droghetti_surf}.
These therefore become de facto organic radicals. Since magnetic coupling between radicals in crystals has been reported in a
large number of studies and different magnetic coupling mechanisms have been proposed \cite{Makarova}, it is interesting to investigate whether any
interaction between Alq$_3^-$ ions is possible. As explained above, the electrons reducing Alq$_3$ molecules to Alq$_3^-$ can be donated by O$_2^-$ impurities in 
La$_{0.7}$Sr$_{0.3}$MnO$_3$/Alq$_3$/AlO$_x$/Co spin-valves. Since most devices have a large charge carrier concentration, it could be that O$_2^-$ impurities
are already incorporated during the fabrication process. On the other hand, at low carriers concentration, impurities are assumed to be driven
into the Alq$_3$ film through a voltage pulse according to the model in Ref. \cite{Bergenti,Riminucci3}. Furthermore, we note that electrons can also be directly 
injected into the Alq$_3$ film by such an applied pulse with no need for O$_2^-$ or other dopant molecules.\\
The exchange coupling between two Alq$_3^-$ ions is calculated by means of the broken symmetry approach similarly to what done for the oxygen
impurities. We highlight that the use of c-DFT is required, not just to force the magnetic moments to be either parallel or anti-parallel, but
also to localize the two added electrons on one molecule each as LSDA would otherwise return the electrons to be shared between the two
molecules in an unphysical way\cite{Kummel}. The results are found not to depend on whether we use the geometry of the neutral molecule or that of
the anion. In fact, the two geometries only differ slightly for the length of the O-Al and N-Al bonds, while, as discussed above, the unpaired electron
of Alq$_3^-$ is localized on the ligands and not along those bonds.\\
$J$ is found to be non-zero and ferromagnetic ($J < 0$) when the ligands of the two ions are stacked in a $\pi -\pi$ configuration as shown in the inset of Fig. \ref{fig.J}. 
Importantly, this relative arrangement of the molecules is also the energetically most favorable one as the overlap between the ions wave-functions is maximized. 
$\vert J\vert$ is equal to $4.5$ meV, which corresponds to about $50$ K, at the calculated equilibrium geometry, for which the distance $d$ between the $\pi$-stacked ligands is $3.35$~\AA. 
$\vert J\vert$ is remarkably large when compared to the typical exchange coupling calculated for radicals \cite{Makarova}. Furthermore, we note
that the considered equilibrium geometry is for a two-ion cluster in vacuum. 
We expect that Alq$_3$ molecules will be more tightly-packed in a device and therefore $d$ will be shorter. $\vert J\vert$  can reach almost room
temperature for $d \approx 2.0$~\AA. This is a quite short distance for
$\pi$-stacked systems, but we note that c-DFT may underestimate $\vert J\vert$ by enforcing a too large localization of the wave functions. 
As such, our results represent conservative estimates. 
The predicted non-zero and eventually large exchange coupling between two Alq$_3^-$ ions is the second important result of this paper.\\
The large $\vert J\vert$-value is because the exchange coupling is direct and not
the result of the kinetic mechanisms often invoked in the case of radicals \cite{Makarova}. The plot of $\vert J\vert$ as a function of $d$ is shown in Fig. \ref{fig.J}. The curve
has a shape similar to the Bethe-Slater curve describing the direct exchange coupling between atomic orbitals of transition metals. 
It presents a maximum at $d = 1.96$~\AA, around which it is well approximated
by a quadratic function. In contrast, at large $d$ ($> 3.5$) \AA, $\vert J\vert$ decays exponentially according to $ce^{-\alpha d}$ with $\alpha = 2.04$~\AA$^{-1}$ and $c = 3.4$ eV.
Such exponential decay at large distances is qualitatively consistent with what assumed by Yu in its original impurity band model. However
he considered the impurities to have hydrogen-like wave functions instead of being $\pi$-stacked molecules and this leads to quantitative variations in the parameters. Differently from Yu's original model, we
proposed in Ref. \cite{Droghetti_model} the exchange coupling to be indirect, rather the direct. The results will be reconsidered based on our new first-principles
predictions in the future.\\
There is an issue that we have neglected so far, namely that having two charged molecules at a close distance is unlikely as they will repel each other. 
Therefore at low carrier concentrations, Alq$_3^-$ ions will be far apart in order to minimize their electrostatic interaction and there will be no magnetic coupling between them. 
Only when the concentration of the electrons is large enough, these will localize on neighbor molecules, which will then be exchanged-coupled. 
In fact, in experiments, the MR is detected only for carriers concentrations of the order of $10^{19}$ cm$^{-3}$ \cite{Riminucci2, Riminucci3}. 
If we model an Alq$_3$ molecule as a sphere of radius $\sim 12$~\AA, we will see that this is a huge concentration, which corresponds to one electron per ten molecules. 
Since each molecule can be surrounded my about nine other molecules, it is unavoidable to find nearest neighbor Alq$_3^-$ ions for such a high electron concentration. \\
Summarizing we suggest that Alq$_3^-$ ions, which are exchange coupled, are present in La$_{0.7}$Sr$_{0.3}$MnO$_3$/Alq$_3$/AlO$_x$/Co spin-valves and 
can carry a spin current in accordance to the impurity band model. We point out that our calculations are for systems in equilibrium and there is no applied electric field. 
The results might therefore be valid only in the very low bias regime, which is anyway when the MR is measured. For large bias voltages, we expect that
electrons will be driven and there might be a strong electron density delocalization as discussed, for example, in Ref. \cite{Navamani}. This means that
dopant electrons will not stay tightly bound to Alq$_3$ forming Alq$_3^-$ ions. This might explain the absence of the MR.\\
Finally we note that spin-transport through Alq$_3$ films was also demonstrated in spin-pumping experiments \cite{SWJiang}. 
The results provided a strong evidence that the spin current was propagated by spin-waves and they were explained by means of the impurity band model by Yu. 
Notably, the estimated carriers concentration in these experiments is similar to that needed to observed the MR in spin-valves. 
Therefore we propose that spin transport involves exchange-coupled Alq$_3^-$ ions also in those spin-pumping experiments.\\

\section{Conclusions}
Based on the results of first-principles calculations, we propose that spin transport in La$_{0.7}$Sr$_{0.3}$MnO$_3$/Alq$_3$/ AlO$_x$/Co spin-valves does not involve oxygen states. 
We then suggest the following alternative picture to explain the MR reported in devices. 
If electrons are injected into the Alq$_3$ film via a voltage pulse, they will reduce some of the Alq$_3$ molecules to Alq$_3^{-}$ ions. 
These are magnetic and, if their concentration is high enough, they will become exchange-coupled and percolate across the film. 
Spin-current is then carried by spin-waves according to the impurity band model.   \\

\section*{Acknowledgment}
This work started at the University of the Basque Country and 
it was sponsored by the ``Ministerio  de  Economia  y  Competitividad'' (MINECO) of Spain through the project ``Transporte Electr\'onico, T\'ermico, y de Espin con la Teor\'ia de Funcionales de Densidad'' (FIS2016-79464-P) and by the Basque Government through the project ``Grupos Consolidados UPV/EHU'' (IT1249-19).
It was concluded at Trinity College Dublin with the economic support provided by the Royal Society and Science Foundation Ireland (SFI).


%\bibliographystyle{cas-model2-names}

\bibliography{cas-refs}


\end{document}